\documentclass[a4paper,12pt]{article}
\renewcommand{\i}{\textrm{i}}
\renewcommand{\d}{\textrm{d}}
\newcommand{\e}{\mathrm{e}}
\begin{document}
\title{Newman-Penrose dyads and the null string motion equations in a 
       background of antisymmetric fields}

\author{K Ilienko\thanks{E-mail address: ilienko@math.ox.ac.uk} \\
        \emph{Balliol College and Mathematical Institute, University
        of Oxford,} \\ \emph{24--29 St Giles', Oxford OX1 3LB, UK}
        \\ \ A A Zheltukhin\thanks{E-mail address:
        zheltukhin@kipt.kharkov.ua} \\ 
        \emph{Kharkiv Institute for Physics and Technology,} \\
        \emph{1 Akademichna Str, Kharkiv 310108, Ukraine}}
\date{}
\maketitle
\begin{abstract}
An application of the Newman-Penrose dyad formalism for description of 
tensionless (null) string dynamics in 4D Minkowski spacetime is studied in
a background of antisymmetric fields. We present a special class of
background antisymmetric fields which admits exact solutions of the
null string motion equations and formulate a
sufficient condition of the absence of interaction of a null string
with such fields. 
\end{abstract}
\newpage
\section{Introduction}
Recently, there has been considerable interest in investigation of the
null string motion equations in different backgrounds of external
fields and in curved spacetimes (e.g. \cite{Veneziano,deVega-Sanchez,Kar1}
and references therein). Finding of the exact solutions for the
equations of motion in such systems is a rather difficult task, mainly
due to the nonlinear character of the motion equations, so, it seems
interesting to study those situations in which these equations possess exact
solutions. It is known that the motion equations for strings in
4D~Minkowski spacetime, null strings in some backgrounds and for
particular types of curved spacetimes can be solved exactly 
\cite{Roshchupkin-Zheltukhin,Kar2,Dabrowski-Larsen}. Being a zero
tension limit of strings \cite{Schild}, null strings possess simpler
equations of motion than those of strings and can be considered
as a zero approximation with the string tension as the perturbation
parameter \cite{Zheltukhin1,deVega-Giannakis-Nicolaidis,Lousto-Sanchez}. 
This explains our current interest in this problem.

Geometrically world sheets of null strings are lightlike (null)
2-surfaces which generalise world lines of massless particles
\cite{Karlhede-Lindstrom,Zheltukhin2,Zheltukhin3}. By convention,
null string interactions with various fields can be analysed into two
types. To the first type we attribute the interactions which violate
the lightlike character of the world sheets and, therefore, lead to
generation of non-zero tension. All other interactions fall into the
second type. In particular, null string interactions with
antisymmetric background tensor fields are of the second type.

As shown in \cite{Zheltukhin4}, 2-component spinors of 3D~Minkowski
spacetime provide particularly convenient framework for studying the general
solutions of null string motion equations in arbitrary antisymmetric
background fields $T_{mn}(x)$. On the other hand, convenience of
adopting 2-spinor description for spacelike and timelike strings in 4D
Minkowski spacetime was demonstrated in
\cite{Hughston-Shaw} and for free null (super) strings and p-branes
in \cite{Bandos-Zheltukhin}. Therefore, it is interesting to investigate the
possibility of application of such a formalism in the form of the
Newman-Penrose dyad calculus in 4D~Minkowski spacetime for the
description of null strings in background antisymmetric fields which
is one of the main goals of the present article. In addition, we find
a special class of background antisymmetric fields which gives rise to exact
solutions of the null string motion equations and formulate a
sufficient condition for the absence of interaction of a null string
with background antisymmetric fields. 

\section{Equations of motion}
Let us consider the action of a null string in an external 
antisymmetric field $T_{mn}(x)$. In the spinor form it can be rewritten as 
follows
\begin{equation}
S = \int \Big[ \rho^\mu \partial_\mu x^{AB'}
o_A\bar{o}_{B'}-
\kappa\varepsilon^{\mu_1 \mu_2}
\partial_{\mu_1} x^{A_1 {B'}_1}
\partial_{\mu_2} x^{A_2 {B'}_2}
T_{A_1 A_2 {B'}_1 {B'}_2}(x^{A B'})\Big]\d^2 \xi ,
\label{2.1}
\end{equation}
where $T_{A_1 A_2 {B'}_1 {B'}_2}$, $x^{A B'}$ and $\d^2 \xi$ 
represent field $T_{mn}$, coordinates of the null string 
$x^m (\tau ,\sigma ) $ and the area element $\d\tau \d\sigma$ of the 
world sheet. We assume that $\xi^{\mu} = (\tau ,\sigma)$ is a
smooth parametrisation of the null-string world sheet, choose  
$\varepsilon^{\tau\sigma} =-\varepsilon^{\sigma\tau}=-1$, 
$\eta_{ij}=\mathrm{diag}(+--\,-)$ and use notation 
$\partial_\mu = \partial / \partial \xi^\mu$. 
The coordinates $\xi$ on the world sheet are dimensionless and we
assume the world sheet vector density $\rho^\mu (\xi)$ to have the
dimensions of inverse length. $\rho^\mu$ ensures invariance of action
(\ref{2.1}) under arbitrary non-degenerate reparametrisations of the
null string world sheet. Interaction constant $\kappa$ has the dimensions of 
inverse length. Spinor fields  $o^A$ and $\iota^A$ form a basis for
2-dimensional complex vector space and obey the
normalisation condition 
\begin{equation}
o_A\iota^A =\overline{(\bar{o}_{B'}\bar{\iota}^{B'})} = \chi ,
\label{2.2}
\end{equation}
where $\chi(x^{AB'})$ represents a possibility of rescaling of the
Newman-Penrose dyad element ${\iota}^A$ at each spacetime point
$x^{AB'}$. The relation  
\begin{equation}
T_{A_1 {B'}_1 A_2 {B'}_2} =
T_{A_1 A_2 {B'}_1 {B'}_2} =
-T_{A_2 A_1 {B'}_2 {B'}_1}
\label{2.3}
\end{equation}
holds for antisymmetric tensor field $T_{mn}$ and we introduce 
$\partial_{AB'}\equiv\partial / \partial x^{AB'}$, 
$\dot x^{AB'}=\partial_\tau x^{AB'}$,
$\acute{x}^{AB'}=\partial_\sigma x^{AB'}$ and
\begin{equation}
3\partial_{[AB'}T_{A_1 A_2 {B'}_1 {B'}_2]}=
\partial_{AB'}T_{A_1 A_2 {B'}_1 {B'}_2}+
\partial_{A_1 {B'}_1}T_{A_2 A {B'}_2 B'}+
\partial_{A_2 {B'}_2}T_{A A_1 B' {B'}_1}.
\label{2.4}
\end{equation}

Variation of action (\ref{2.1}) results in the equations describing the 
null string dynamics
\begin{equation}
\begin{array}{l}
\rho^\mu \partial_\mu x^{AB'} o_A =0, \\
\partial_{\mu} x^{AB'}o_A\bar{o}_{B'} = 0, \\
\partial_\mu (\rho^\mu o_A\bar{o}_{B'})+   
3\kappa\varepsilon^{\mu_1 \mu_2}
\partial_{\mu_1} x^{A_1 {B'}_1}
\partial_{\mu_2} x^{A_2 {B'}_2}
\partial_{[AB'}T_{A_1 A_2 {B'}_1 {B'}_2]}=0, 
\end{array}
\label{2.5}
\end{equation}
and complex conjugate of them.
The first equation in (\ref{2.5}) and its complex conjugate imply 
\begin{equation}
\rho^\mu\partial_\mu x^{AB'} = e o^A \bar{o}^{B'},
\label{2.6}
\end{equation}
where $e(\xi)$ is an arbitrary real-valued function with
transformation properties of a scalar world sheet density. Assuming that
$\rho^\tau$ is a nowhere zero function we rewrite this equation as 
\begin{equation}
\dot{x}^{AB'}=
\frac{e}{\rho^\tau} o^A \bar{o}^{B'}-
\frac{\rho^\sigma}{\rho^\tau}\acute{x}^{AB'}.
\label{2.7}
\end{equation}
Taking into account the second equation in (\ref{2.5}) we obtain 
\begin{equation}
\acute{x}^{AB'}o_A\bar{o}_{B'}=0.
\label{2.8}
\end{equation}
This equation yields representation for spin-tensor
$\acute{x}^{AB'}$ in the form\footnote{Representations (\ref{2.7}) and
(\ref{2.9}) imply $\dot{x}^2 = (\rho^\sigma /\rho^\tau )^2
\acute{x}^2$ and $\dot{x}\acute{x}= -(\rho^\sigma
/\rho^\tau)\acute{x}^2$. Hence, the determinant of the induced metric
on the null string world sheet vanishes identically
($\dot{x}^2\acute{x}^2 - (\dot{x}\acute{x})^2 = 0$). This verifies
that the action principle (\ref{2.1}) provides a description of a null
string world sheet.}
\begin{equation}
\acute{x}^{AB'}= o^A\bar{r}^{B'}+r^A\bar{o}^{B'},
\label{2.9}
\end{equation}
where condition $o_A r^A \not =0$ is imposed on the spinor field $r^A$.

Action (\ref{2.1}) is invariant with respect to the following transformations
\begin{equation}
\tilde{o}^A = o^A, \,\,\,\,
\tilde{\iota}^A = {\iota}^A + uo^{A}, 
\label{2.10}
\end{equation}
and
\begin{equation}
\tilde{o}^A = \textrm{e}^v o^{A}, \,\,\,\,
\tilde{\iota}^A = \textrm{e}^{-v} \iota^{A}, \,\,\,\,
\tilde{\rho}^\mu = \textrm{e}^{-(v+\bar{v})}\rho^\mu
\label{2.11}
\end{equation}
with complex-valued functions $u(\xi)$ and $v(\xi)$. Expanding spinor
field $r^A$ in the basis ($o^A$,~$\iota^A$) as $r^A = p o^A + q\iota^A$,
where functions $p(\xi)$ and $q(\xi)$ are complex-valued, 
representing $q$ in the form $q=|q|\exp(\i\varphi)$ and using
(\ref{2.9}) we get
\begin{equation}
\acute{x}^{AB'}=
 (p+\bar{p}) o^A\bar{o}^{B'} +
 |q|(\textrm{e}^{-\i\varphi} o^A\bar{\iota}^{B'} +
\textrm{e}^{\i\varphi}\iota^A\bar{o}^{B'}).
\label{2.12}
\end{equation}
Carrying out successive transformations (\ref{2.10}) and (\ref{2.11})
with parameters \\ $u= |q|^{-1}\bar{p}\exp(-\i\varphi )$ and
$v=-[\ln (\rho^\tau /e ) + \i\varphi ]/2$ we obtain
\begin{equation}
\dot{x}^{AB'} = o^A \bar{o}^{B'} 
- \frac{\rho ^\sigma}{\rho ^\tau}\acute{x}^{AB'}, \,\,\,\,
\acute{x}^{AB'} = |q|(o^A \bar{\iota}^{B'} + \iota^A \bar{o}^{B'}).
\label{2.13}
\end{equation}
Here we omitted the tildes over the basis spinors, because motion
equations (\ref{2.5}) are invariant with respect to transformations
(\ref{2.10}) and (\ref{2.11}). Using the possibility of rescaling the
dyad element $\iota ^A$
\begin{equation}
\tilde{o}^A = o^{A}, \,\,\,\,
\tilde{\iota}^A = \lambda\iota ^{A}, \,\,\,\,
\tilde{\chi} = \lambda\chi,
\label{2.14}
\end{equation}
we incorporate factor $|q|$ into $\iota ^A$ and
write (\ref{2.13}) in the form 
\begin{equation}
\dot{x}^{AB'} = o^A \bar{o}^{B'} 
- \frac{\rho ^\sigma}{\rho ^\tau}\acute{x}^{AB'}, \,\,\,\,
\acute{x}^{AB'} = o^A \bar{\iota}^{B'} + \iota^A \bar{o}^{B'}.
\label{2.15}
\end{equation}
Using (\ref{2.15}) we represent the last equation in (\ref{2.5}) as
\begin{equation}
\partial_\mu (\rho^\mu o_A\bar{o}_{B'}) =
\i\kappa (\chi Q_{AD'}\bar{o}^{D'}\bar{o}_{B'} - 
\bar{\chi}o_A Q_{CB'}o^C),
\label{2.16}
\end{equation}
where spin-tensor $Q_{AB'}$ corresponds to vector 
$Q^k = \varepsilon^{klmn}\partial_{[l}T_{mn]}$ with  
$3!\partial_{[l}T_{mn]}=
\varepsilon_{lmnp}\varepsilon^{pqrs}\partial_{q} T_{rs}$
and we take 
$\varepsilon_{0123} = - \varepsilon^{0123} = 1$.

Invariance of action (\ref{2.1}) under 
arbitrary reparametrisations of the world sheet implies (via the
second Noether's theorem \cite{Noether,Barbashov-Nesterenko}) 
that the solutions of the motion equations 
depend upon two arbitrary real-valued functions. We fix one of them
using a condition 
\begin{equation}
\rho^\sigma =0.
\label{2.17}
\end{equation}
Then, the null string motion equations take the form
\begin{equation}
\begin{array}{l}
\dot{x}^{AB'} = o^A \bar{o}^{B'}, \,\,\,\,
\acute{x}^{AB'}= o^A \bar{\iota}^{B'} + \iota^A \bar{o}^{B'}, \\
(\rho^\tau o_A\bar{o}_{B'})^. =
\i\kappa (\chi Q_{AD'}\bar{o}^{D'}\bar{o}_{B'} - 
\bar{\chi}o_A o^C Q_{CB'}),
\end{array}
\label{2.18}
\end{equation}
and the first two of them result in the Virasoro constraints
\begin{equation}
\dot{x}^2=0 \quad\mbox{and}\quad\dot{x}\acute{x} =0
\end{equation}
\label{2.19}
appearing in the standard theory of null strings.

\section{Absence of interaction with antisymmetric field}
Projecting equation (\ref{2.16}) on the spin-tensor basis element
$o^A\bar{\iota}^{B'}$ we obtain
\begin{equation}
o^A \rho^\mu \partial_\mu o_A =
\i\kappa \chi o^A Q_{AB'}\bar{o}^{B'}.
\label{3.0}
\end{equation}
When the external antisymmetric field vanishes $Q_{AB'}$ is equal to 
zero and (\ref{3.0}) gives a reparametrization invariant equation of a
null string world sheet which is generated by null geodesics
(cf. \cite{Schild}). Thus, setting $Q_{AB'}$ to zero we get under gauge
condition (\ref{2.17})  equation 
\begin{equation}
\dot{o}^A o_A = 0
\label{3.1}
\end{equation} 
as a sufficient condition for a null string
world sheet to be a geodesic one. We note that this condition, together
with the null string motion equations and the Virasoro constraints, is
invariant under non-degenerate transformations of the form
(see \cite{Zheltukhin4})  
\begin{equation}
\tilde{\tau} = \tilde{\tau}(\tau,\sigma)\quad\mbox{and}
\quad\tilde{\sigma} = \tilde{\sigma}(\sigma).
\label{3.2}
\end{equation}

Equation (\ref{3.1}) under gauge condition (\ref{2.17}) shows 
that the sufficient condition of the absence of
the null string interaction with an antisymmetric external field is
\begin{equation}
Q_{AB'} o^A \bar{o}^{B'} = 0,
\label{3.3}
\end{equation}
or $Q^k \dot{x}_k = 0$ in vector form. Since vector $Q^k$ is dual to
the field strength tensor $\partial_{[l}T_{mn]}$ we observe that
condition (\ref{3.3}) is equivalent to the impossibility of
constructing a non-zero 4-volume out of the field strength tensor and
vector $\dot{x}^k$. Thus, the measure of this volume can be taken as
a measure of interaction.

This case corresponds to system (\ref{A.8}) with $\phi = 0$
(cf. equation (\ref{3.1})). The solution for spinors $o^A$ and
$\iota^A$ is given by
\begin{eqnarray}
o^A  & = & \alpha^A (\sigma)\e ^{\frac{m}{2}}, 
\nonumber \\
\iota^A & = & 
\chi\big[\beta^A (\sigma) + \int_{o}^{\tau} (\acute{o}^A + \i\mu o^A)
\e ^{\frac{\overline{m}}{2}}\d\tau\big]\e ^{-\frac{m}{2}}
\label{3.4}
\end{eqnarray}
and (\ref{2.2}) implies
\begin{equation}
\alpha_A\beta^A + \alpha_A \acute{\alpha}^A R = 
\chi \e ^{-\frac{m-\overline{m}}{2}},
\label{3.5}
\end{equation}
where
\begin{equation}
R = \int_{0}^{\tau}\e ^{\frac{m +
\overline{m}}{2}}\d\tau ,
\label{3.6}
\end{equation}
as the normalisation condition. Representation (\ref{3.4}) allows to
write (\ref{2.18}) in the form
\begin{equation}
\begin{array}{l}
\dot{x}^{AB'} = \alpha^A\bar{\alpha}^{B'}\dot{R}, \\
\acute{x}^{AB'} = \alpha^A\bar{\beta}^{B'} + 
                          \beta^A\bar{\alpha}^{B'} +
                          (\alpha^A\bar{\alpha}^{B'}R)', \\
\rho^{\tau}\ddot{R} + \dot{\rho^{\tau}}\dot{R} = 
\i\kappa\dot{R}[(\beta^A + 
\acute{\alpha}^A R)Q_{AB'}\bar{\alpha}^{B'} - 
\alpha^A Q_{AB'}(\bar{\beta}^{B'} + \acute{\bar{\alpha}}^{B'} R)].
\end{array}
\label{3.7}
\end{equation}
Equation (\ref{3.3}) and the first two equations in (\ref{3.7}) ensure that
the restriction of the background antisymmetric field $T_{mn}(x)$ to
the null string world sheet given by spin-tensor $Q_{AB'}$ depends on
$\tau$ only through $R$. Hence making use of (\ref{3.4}) in
(\ref{3.3}) we obtain 
\begin{equation}
Q_{AB'} = \alpha_A\bar{w}_{B'}(R,\sigma) + w_{A}(R,\sigma)\bar{\alpha}_{B'}. 
\label{3.8}
\end{equation}
Imposing the second gauge condition in a form
\begin{equation}
\rho^\tau = 1,
\label{3.9}
\end{equation}
analogues to \cite{Zheltukhin4}, we are able to find solution for $R$
of the last equation in system (\ref{3.7}) in implicit form 
\begin{equation}
\tau = \int_{0}^{R} \bigg[ \int_{0}^{y} F(x,\sigma )\d x \bigg] ^{-1}
\d y ,
\label{3.10}
\end{equation}
where the function $F(x,\sigma)$ is given by 
\begin{equation}
F(x,\sigma ) = \i\kappa [
(\alpha_A\beta^A + x\alpha_A\acute{\alpha}^A )
\bar{\alpha}^{B'}\bar{w}_{B'}(x,\sigma)
- \alpha^A w_{A}(x,\sigma)
(\bar{\alpha}_{B'}\bar{\beta}^{B'} + 
x\bar{\alpha}_{B'}\acute{\bar{\alpha}}^{B'})].
\label{3.11}
\end{equation}

Taking into account invariance of the null string motion equations
under transformations (\ref{3.2}) we may choose new local coordinates by
$\tilde{\tau} = R(\tau,\sigma)$ and $\tilde{\sigma} = \sigma$. Then,
the first equation in system (\ref{3.7}) results in the equation of
motion for a free null string 
\begin{equation}
\ddot{\tilde{x}}^{AB'} (\tilde{\tau},\tilde{\sigma}) = 0
\label{3.12}
\end{equation}
(or $\ddot{\tilde{x}}^m = 0$ in the vector form), where the
differentiation is performed with respect to $\tilde{\tau}$. This
equation possesses exact solution and can be easily integrated, thus
we find that the null string motion equations in background
antisymmetric fields which obey conditions (\ref{3.3}) can also be
solved exactly.

\section{Conclusion}
In this paper null string motion equations in arbitrary background of
antisymmetric fields are considered. We find a special class of the 
fields, singled out by condition (\ref{3.3}), which can be eliminated
by a particular choice of parametrisation for the null string world
sheet, such that the null string motion equations are reduced to those of a
free null string and, therefore, admit exact solution. At this point,
it seems interesting to generalise the above mentioned results to the
case of tensionless p-branes in antisymmetric external fields and for
d-dimensional spacetimes. 

\section{Acknowledgements}
This work is supported, in part, by Dutch Government and INTAS Grants
No~94--2317, No~93-633-ext and No~93-127-ext. The first author 
would like to acknowledge the Dervorguilla Graduate Scholarship
awarded by Balliol College. He is also grateful to Profs~R~Penrose and
Yu~P~Stepanovsky for fruitful discussions. The second author wish to
acknowledge financial support by the Government Fund for Fundamental
Research of the Ministry of Research and Technology of Ukraine and a
High Energy Physics Grant awarded by the ministry. 

After submitting this article the authors learnt about recent paper
\cite{Jensen-Lindstrom} which contains an interesting approach to the
description of interaction of tensionless strings with other fields
and has an interesting intersection with our consideration. We are
grateful to Dr B~Jensen who informed us about this work.

\appendix
\newpage
\section{Analysis of consistency relations}
Let us analyse consistency relations for the representations of
$\dot{x}^{AB'}$ and $\acute{x}^{AB'}$. Since
$(\dot{x}^{AB'})'$ is equal to $(\acute{x}^{AB'})^.$ the spinors
$o^A$ and $\iota^A$ obey the relation
\begin{equation}
\acute{o}^A\bar{o}^{B'} + o^A\acute{\bar{o}}^{B'} =
o^A\dot{\bar{\iota}}^{B'} + \dot{o}^A\bar{\iota}^{B'} +
\iota^A\dot{\bar{o}}^{B'} + \dot{\iota}^A\bar{o}^{B'}.
\label{A.1}
\end{equation}
Multiplying both sides of equation~(\ref{A.1}) by
$o_A\bar{o}_{B'}$ we get
\begin{equation}
\bar{\chi}\dot{o}^A o_A + \chi\dot{\bar{o}}^{B'}\bar{o}_{B'} = 0,
\label{A.2}
\end{equation}
which yields
\begin{equation}
\dot{o}^A o_A + \i\chi\phi = 0,
\label{A.3}
\end{equation}
where $\phi (\xi)$ is an arbitrary real-valued function. Allowing for
(\ref{2.2}), one can write
\begin{equation}
\dot{o}^A = \frac{1}{2}\dot{m}o^A - \i\phi\iota^A.
\label{A.4}
\end{equation}
Here $m(\xi)$ is an arbitrary complex-valued function. Substituting
(\ref{A.4}) into (\ref{A.1}) and projecting the obtained equation on
$o_A\bar{\iota}_{B'}$ we find
\begin{equation}
(\dot{\iota}^A + \frac{1}{2}\dot{\overline{m}}\iota^A -
\acute{o}^A )o_A = 0.
\label{A.5}
\end{equation}
From (\ref{A.5}) it follows that
\begin{equation}
\dot{\iota}^A + \frac{1}{2}\dot{\overline{m}}\iota^A - \acute{o}^A  = 
\tilde{\mu} o^A.
\label{A.6}
\end{equation}
Substituting again (\ref{A.4}) and (\ref{A.6}) into (\ref{A.1}) we
obtain
\begin{equation}
\tilde{\mu} = \i\mu,
\label{A.7}
\end{equation}
where $\mu (\xi)$ is an arbitrary real-valued function. Thus, consistency
relation (\ref{A.1}) results in a system of equations for the basis spinors
$o^A$ and $\iota^A$
\begin{equation}
\begin{array}{l}
\dot{o}^A = \frac{1}{2}\dot{m} o^A - \i\phi\iota^A , \\
\dot{\iota}^A = \i\mu o^A - \frac{1}{2}\dot{\overline{m}}\iota^A 
+ \acute{o}^{A}.
\end{array}
\label{A.8}
\end{equation}

\newpage

\end{document}